\documentclass[aps,prl,reprint,amsmath,amssymb,superscriptaddress]{revtex4-2}

\usepackage[utf8]{inputenc}
\usepackage[T1]{fontenc}
\usepackage{graphicx}
\usepackage{xcolor}
\usepackage[normalem]{ulem}

\AtBeginDocument{%
  \setcounter{dbltopnumber}{1}%
}
\usepackage{bm}

\usepackage{booktabs}
\usepackage{afterpage}

\newcommand{\pR}{\rho_R}
\newcommand{\Cov}{\mathrm{Cov}}
\newcommand{\E}{\mathbb{E}}
\makeatletter
\AtBeginDocument{%
  \let\smthesubsection\thesubsection
  \let\smthesubsubsection\thesubsubsection
  \let\smtheeqprefix\theequation@prefix
  \let\smhangfromsection\@hangfrom@section
  \let\smsectioncntformat\@sectioncntformat
  \let\smmaketitle\maketitle
  \let\smauthor\author
  \let\smaffiliation\affiliation
  \let\smemail\email
  \let\smthanks\thanks
  \let\smAFjoin\@AF@join
}
\newcommand{\smskiplabel}[1]{\sm@nextlabel{#1}{}}
\newcommand{\smrelabel}[2]{\sm@nextlabel{#1}{#2}}
\newcommand{\sm@nextlabel}[2]{%
  \def\smwanted{#1}\def\smrename{#2}%
  \let\smoldlabel\label
  \def\label##1{%
    \let\label\smoldlabel
    \def\smlabelarg{##1}%
    \ifx\smlabelarg\smwanted
      \ifx\smrename\@empty\else\smoldlabel{#2}\fi
    \else\smoldlabel{##1}\fi}%
}
\newcommand{\smfrontmatter}{%
  \smskiplabel{FirstPage}%
  \let\maketitle\smmaketitle
  \let\author\smauthor
  \let\affiliation\smaffiliation
  \let\email\smemail
  \let\thanks\smthanks
  \let\@AF@join\smAFjoin
  \let\@AAC@list\@empty \let\@AFF@list\@empty \let\@AFG@list\@empty
  \let\@address\@empty
  \frontmatter@init          %
}
\makeatother

\begin{document}

\title{Bypassing the Entropy Production Speed Limit of State-to-State Transitions with Adaptive Resetting}

\author{Daniel B. K. Chu}
\affiliation{School of Chemistry, Tel Aviv University, Tel Aviv 6997801, Israel}

\author{R\'emi Goerlich}
\affiliation{School of Chemistry, Tel Aviv University, Tel Aviv 6997801, Israel}

\author{Itamar Steinitz}
\affiliation{School of Chemistry, Tel Aviv University, Tel Aviv 6997801, Israel}

\author{Yael Roichman}
\email{roichman@tauex.tau.ac.il}
\affiliation{School of Chemistry, Tel Aviv University, Tel Aviv 6997801, Israel}
\affiliation{School of Physics, Tel Aviv University, Tel Aviv 6997801, Israel}
\affiliation{The Center for Physics and Chemistry of Living Systems, Tel Aviv University, Tel Aviv 6997801, Israel.}

\author{Barak Hirshberg}
\email{hirshb@tauex.tau.ac.il}
\affiliation{School of Chemistry, Tel Aviv University, Tel Aviv 6997801, Israel}
\affiliation{The Center for Physics and Chemistry of Living Systems, Tel Aviv University, Tel Aviv 6997801, Israel.}
\affiliation{The Center for Computational Molecular and Materials Science, Tel Aviv University, Tel Aviv 6997801, Israel.}

\date{\today}

\begin{abstract}
State-to-state transitions in colloids and cold atoms (trap compression, bit erasure) are often achieved through driving, whose speed and entropy production are constrained by a fundamental bound. We show that this bound can be surpassed by changing the transport mechanism. Adaptive resetting enables transitions through nonlocal jumps, with rates learned from reset-free trajectories. A hybrid protocol, combining resetting with a switch to the target potential, surpasses the diffusive bound in three transformations. Finally, we derive a speed limit governing coupled diffusion-reset dynamics.
\end{abstract}

\maketitle

Transforming a system from one state to another in finite time is a task of broad interest: a colloidal particle or cold atom driven between two trap configurations~\cite{martinez2016,schaff2010}, a bit of information erased by collapsing a double well into a single one~\cite{berut2012}, or a molecule prepared at a transition state~\cite{debnath2019}. The simplest recipe works for a target that is the equilibrium state of some potential: abruptly switch the potential from $U_0$ to $U_{\rm tgt}$ and wait for thermal relaxation. But this generality comes at a cost. The slowest mode of the new potential sets the relaxation time, while the abrupt switch performs excess work that is dissipated during the subsequent relaxation.

Driving, i.e., shaping $U(x,t)$ in time, can improve both speed and thermodynamic efficiency. Engineered swift equilibration (ESE) constructs smooth protocols that land exactly on the target equilibrium in a chosen time~\cite{martinez2016}. Finite-time optimal control also minimizes dissipation over a family of driving potentials~\cite{schmiedl2007}, but the end-of-drive state need not be an equilibrium. The target is then attained only after an additional thermal relaxation. Optimal-transport theory identifies the minimum-dissipation protocol that reaches the prescribed state in finite time~\cite{aurell2011,aurell2012,nakazato2021}. All driving protocols, however, share one feature: the probability density evolves under a continuity equation, carried by the velocity field that the potential produces. From this follows a universal lower bound~\cite{aurell2011,aurell2012,nakazato2021},
\begin{equation}
\Sigma \;\ge\; \frac{W_2^2\!\left(\rho_0,\rho_\tau\right)}{D\,\tau},
\label{eq:drivingbound}
\end{equation}
limiting the entropy production $\Sigma$ of \emph{any} purely diffusive drive that carries an initial density $\rho_0$ to $\rho_\tau$ in a time $\tau$ by the squared Wasserstein distance between the endpoints. Faster preparation costs more, with a $1/\tau$ scaling that no ingenuity in protocol design can evade. Here, we ask whether this limit can be changed not by optimizing the driving protocol, but by changing the transport mechanism itself.

Stochastic resetting has recently emerged as a powerful tool for controlling state-to-state transitions~\cite{goerlich2024,goerlich2026}. Rather than deforming the potential, we repeatedly reset the system to an independent sample of the reset distribution.
This process can relax faster than thermal equilibration, but usually toward a different destination: its own non-equilibrium steady state (NESS)~\cite{sherf2026}. A resetting-based swift-equilibration protocol has been demonstrated~\cite{goerlich2024}, but only when its NESS happened to coincide with the target equilibrium state. This coincidence is not widespread, which kept resetting-based state preparation to a few isolated examples. 

The first result of this Letter is that \emph{adaptive} resetting largely removes this restriction: a state-dependent resetting probability $p_\theta(x)$, parameterized by a neural network and trained within the framework of Ref.~\cite{keidar2025}, can steer the resetting NESS onto a prescribed target whenever the reset distribution can sufficiently replenish the regions from which the underlying dynamics drains probability (see Sec.~\ref{sm:realiz} of the Supplemental Material (SM)~\cite{sm}). When exact realization is not possible, resetting can still bring the system close to the target before subsequent relaxation in the target potential. Together with a final relaxation step, adaptive resetting provides a general route to state preparation even when the target itself is not realizable as a resetting NESS.

\begin{figure*}[t]
\includegraphics[width=0.70\textwidth]{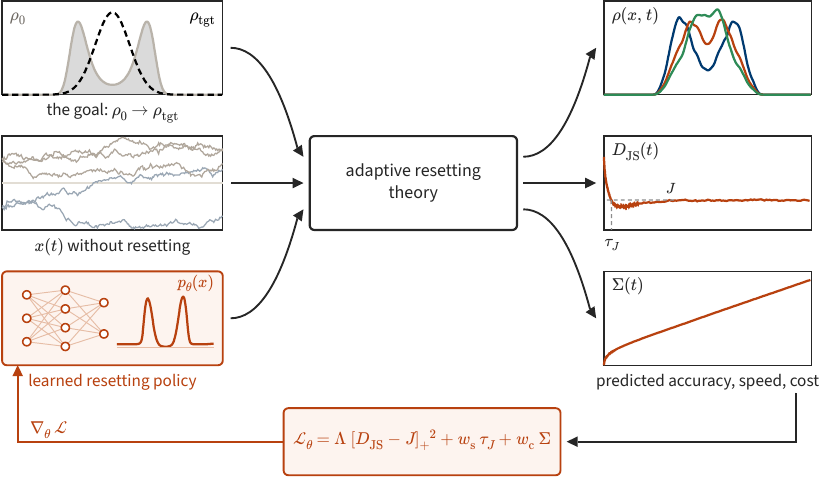}
\caption{Illustration of the framework, which learns a resetting policy $p_\theta(x)$ that brings an initial state $\rho_0$ to a target $\rho_{\rm tgt}$ (here, double well to transition state). An ensemble evolved in $U_0$ \emph{without} resetting and the learned policy $p_\theta(x)$ feed the renewal propagator of Eq.~\eqref{eq:renewal}, which predicts the transient density $\rho(x,t)$, its distance to the target $D_{\rm JS}(t)$, and the entropy production $\Sigma(t)$. The loss, Eq.~\eqref{eq:loss}, is computed from these, and its gradient $\nabla_\theta\mathcal{L}$ updates $p_\theta$. No resetting is simulated during training.}
\label{fig:method}
\end{figure*}

Our second main result concerns the thermodynamic cost. Because a reset event induces a discontinuous jump in probability, it introduces source and sink terms to the Fokker--Planck equation rather than continuous probability flux. Consequently, the continuity equation underlying Eq.~\eqref{eq:drivingbound} no longer applies. We exploit this nonlocality using a \emph{hybrid protocol}: adaptive resetting first achieves rapid probability redistribution, after which switching to the target potential completes state preparation via relaxation. Analogous to optimal-control schemes that pair driven transport with final relaxation, this two-step architecture achieves a prescribed target accuracy and, for every transformation we study, can do so at a cost below the driving bound of Eq.~\eqref{eq:drivingbound}, accessing a regime unavailable to purely diffusive driving.

This does not violate the second law: the entropy production $\Sigma$ of the coupled diffusion-reset dynamics incorporates both diffusive dissipation~\cite{sekimoto2010,seifert2012} and the entropy production of the reset-jump channel~\cite{fuchs2016,mori2023}. Resetting is thus not cost-free; instead, it opens a nonlocal transport channel that is not subject to the diffusive driving bound of Eq.~\eqref{eq:drivingbound}. We show that the coupled dynamics obey a different speed limit, derived below for coupled diffusion-reset transport at fixed reset mobility (see Eq.~\eqref{eq:floor} and the End Matter). 

We demonstrate the hybrid protocol in three state-preparation
tasks: harmonic compression, a distribution
localized at the transition state of a double well, and bit
erasure. In each case, the reference protocol is an abrupt
switch from the initial potential $U_0$ to the target potential
$U_{\rm tgt}$, followed by thermal relaxation. For bit erasure,
the target is a single-well centered at one minima of the double well, following the standard
colloidal realization of Landauer erasure.
We quantify the preparation time, accuracy, and entropy
production for each transformation. With suitable resetting
policies and handoff times, all three transformations cross the
bound for purely diffusive driving. The advantage is largest
and most robust for bit erasure, while transition-state
preparation crosses the bound only within a narrow parameter
range.

We consider a Brownian particle in one dimension, evolving under overdamped Langevin dynamics in a \emph{static} potential $U_0(x)$ with diffusion coefficient $D$ and inverse temperature $\beta$ (we set $k_BT=\gamma=D=1$). The system starts in the Boltzmann state $\rho_0\propto e^{-\beta U_0}$, which also serves as the reset distribution, $\rho_R=\rho_0$. We treat $\rho_R$ as an available reset reservoir; the cost of preparing samples from this distribution is not included in the effective entropy production of the diffusion--reset dynamics. The trajectories were integrated using Euler--Maruyama with time step $\Delta t$. Under adaptive resetting the particle diffuses in $U_0$ and is teleported to an independent sample of $\rho_R$ with a \emph{state-dependent} probability $p_\theta(x)\in(0,1)$ at each step, or, equivalently, at a position-dependent resetting rate $r(x)=-\ln[1-p_\theta(x)]/\Delta t$. We parameterize the policy $p_\theta$ by a feed-forward neural network with parameters $\theta$ and a sigmoid output (see Sec.~\ref{sm:training}).

The key practical feature of the adaptive resetting framework~\cite{keidar2025} is that the full time-dependent density under resetting can be predicted from an ensemble of $N$ trajectories, each evolved for $L$ steps in $U_0$ \emph{without} resetting. Writing $X^i_k$ for the position of trajectory $i$ at step $k$ (time $k\,\Delta t$), and $\Psi^i_{n}=\prod_{k=1}^{n-1}\bigl[1-p_\theta(X^i_k)\bigr]$ for the probability that trajectory $i$ has survived to step ${n}$ without being reset, the density $\rho_n(x)\equiv\rho(x,n\Delta t)$ of the resetting process obeys the renewal relation
\begin{equation}
\begin{split}
\rho_{n}(x) ={}& \frac{1}{N}\sum_{i=1}^{N} \Psi^i_{n}\,\delta(x-X^i_{n})\\
&+ \sum_{k=1}^{n-1} \Bigl(\tfrac{1}{N}\sum_{i=1}^{N} \Psi^i_k\,p_\theta(X^i_k)\Bigr)\,
\rho_{n-k}(x),
\end{split}
\label{eq:renewal}
\end{equation}
which is differentiable in $\theta$. A single stored ensemble thus yields, for \emph{any} candidate resetting policy $p_\theta$, the entire transient $\rho_1,\dots,\rho_{L}$ and its NESS limit.

\begin{figure*}[t]
\includegraphics[width=\textwidth]{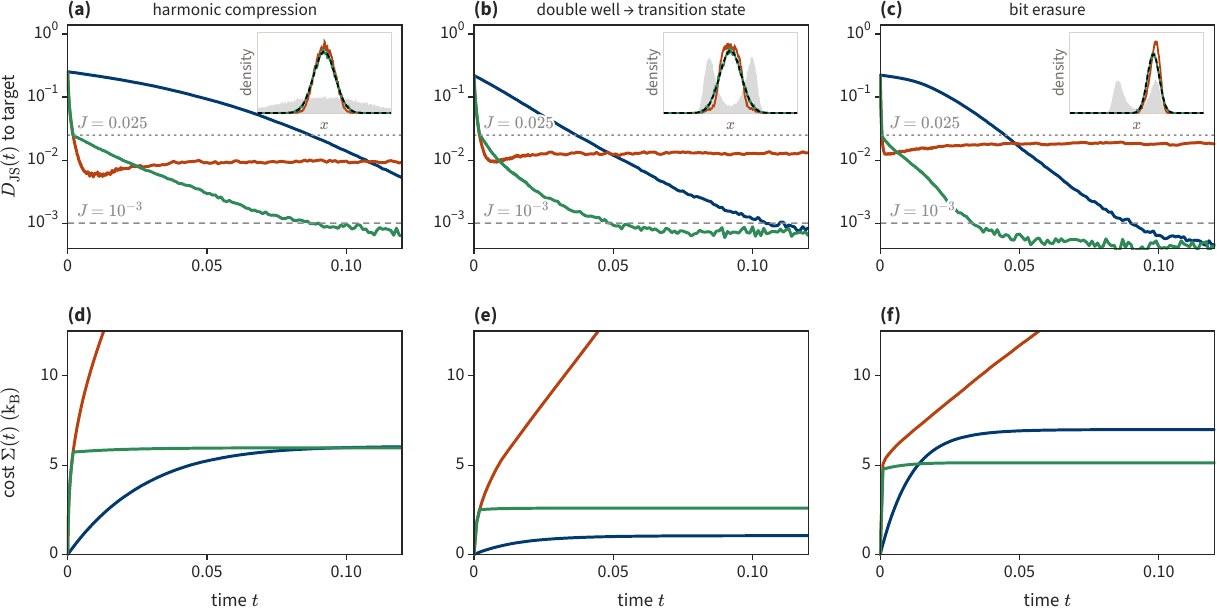}
\caption{Each column is a transformation: harmonic compression, double well to transition state, and bit erasure. Each color is a different protocol: the potential switch (blue), a representative trained adaptive resetting policy (orange), and the hybrid (green), which hands off to the switch once resetting has reached the coarse accuracy. Panels (a)--(c) show the distance to the target over time. For compression, the transition state, and erasure, resetting reaches the coarse accuracy $J=0.025$ (dotted) $45$, $19$, and $69$ times faster than the switch, respectively, then stalls. The hybrid reaches the tight accuracy $J=10^{-3}$ (dashed). The insets show the resulting steady states: the initial $\rho_0$ (grey, also the reset distribution), the target (dashed), the resetting NESS (orange), and the hybrid's final state (green). Panels (d)--(f) show the entropy production: the switch and the hybrid saturate once relaxed, while pure resetting requires persistent entropy production to hold its NESS.}
\label{fig:three}
\end{figure*}

\begin{figure*}[t]
\includegraphics[width=\textwidth]{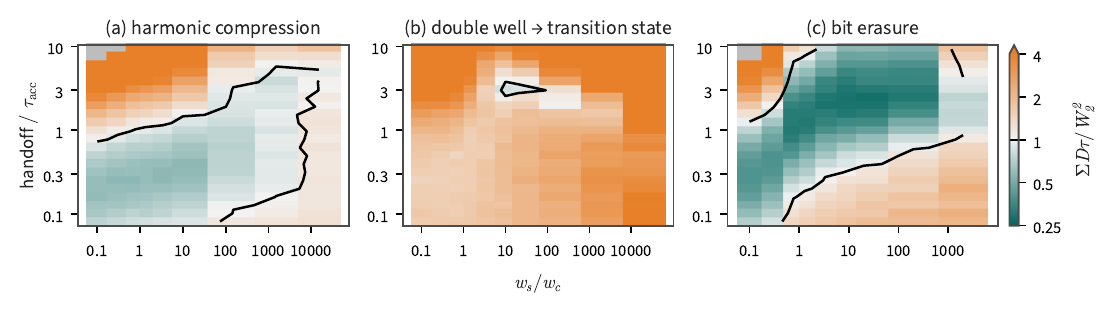}
\caption{
Thermodynamic advantage of hybrid protocols for (a)~harmonic compression, (b)~double-well transition-state preparation, and (c)~bit erasure.
The normalized entropy production $\Sigma D\tau/W_2^2(\rho_0,\rho_\tau)$ is plotted against the optimization weight ratio $w_s/w_c$ and the handoff time in units of $\tau_{\rm acc}$, the time for resetting alone to reach $J=0.025$.
Each weight ratio represents a separately trained resetting policy evaluated across multiple handoff times.
Green denotes entropy production below the purely diffusive driving bound, Eq.~\eqref{eq:drivingbound}, orange denotes values above it, and the black contour marks equality.
All values correspond to the tight accuracy $J=10^{-3}$, with $W_2$ evaluated using the actual density reached at $\tau$.
The color scale saturates at $4$ (maximum value $37.5$), and gray cells indicate conditions that did not reach the prescribed accuracy within the simulation time.
}
\label{fig:bounds}
\end{figure*}

To train $p_\theta$ we define a loss that combines three terms: the distance from the target, the relaxation speed, and the thermodynamic cost. We impose accuracy as a \emph{constraint} and weight speed and cost:
\begin{equation}
\mathcal{L}(\theta) \;=\;
\Lambda\,\bigl[\overline{D}_{\rm JS}-J\bigr]_+^{\,2}
\;+\; w_s\,\hat\tau_J
\;+\; w_c\,\hat\Sigma .
\label{eq:loss}
\end{equation}
In Eq.~\eqref{eq:loss}, $\Lambda$ is the (large) penalty weight enforcing the accuracy constraint; $w_s,w_c$ then trade speed against cost at fixed accuracy.
Here $\overline{D}_{\rm JS}$ is the propagator's late-time (steady-state) Jensen--Shannon divergence to the target and $[\,\cdot\,]_+=\max(\cdot,0)$, taken before the square so the penalty is one-sided, i.e., nonzero only when $\overline{D}_{\rm JS}>J$; a bare square would also penalize over-accuracy, so the first term is a penalty that vanishes once the steady state is within the prescribed accuracy $J$. The speed term $\hat\tau_J=\sum_{n=1}^{L} \sigma\!\left[(D_{\rm JS}(\rho_n,\rho_{\rm tgt})-J)/\epsilon\right]\Delta t$, with $\sigma$ a logistic sigmoid and $\epsilon$ a smoothing width, is a differentiable proxy for the time to reach $J$. Finally, $\hat\Sigma$ is the entropy production accumulated over the dynamics. The entropy-production rate of the resetting process is
\begin{equation}
\dot\Sigma \;=\; \int \frac{j^2}{D\,\rho}\,dx
\;+\; \mathrm{Cov}_{\rho_R}\!\left(h,\ln h\right),
\label{eq:ep}
\end{equation}
with the probability current $j=-D(\partial_x\rho+\beta\rho\,\partial_xU_0)$ and $h=r\rho/\rho_R$.
The first term is the standard diffusive entropy-production rate, while the second is the Schnakenberg entropy production of the reset-jump channel~\cite{schnakenberg1976} (See Sec.~\ref{sm:ep}). For constant $r$, it reduces to $r[D_{\rm KL}(\rho\|\rho_R)+D_{\rm KL}(\rho_R\|\rho)]$, the symmetrized relative-entropy form of Ref.~\cite{mori2023}. The relation to the alternative steady-state decomposition of Ref.~\cite{mori2023} is detailed in Sec.~\ref{Sec:SM_ep} of the SM. 
Evaluated on $\rho_n$ and accumulated in time (Sec.~\ref{sm:cost}), Eq.~\eqref{eq:ep} makes the cost differentiable in $\theta$ alongside the distance and the speed proxy. We make each term dimensionless: we divide the accuracy penalty by $J^2$, and the speed and cost by their values at the first training epoch. We train the resetting policies of Figs.~\ref{fig:three} and \ref{fig:bounds} with the weights listed in Sec.~\ref{sm:training}. Finally, note that Eq.~\ref{eq:ep} is the entropy production of the effective diffusion--reset dynamics; it does not include the implementation-dependent work required to realize the physical reset mechanism.

Figure~\ref{fig:method} illustrates the framework. We first run an ensemble of trajectories in $U_0$ \emph{without} resetting. We initialize the network with random parameters and feed it, with the trajectories, to the renewal propagator, which predicts the density, its distance to the target, and the cost. The loss and its gradient then update the network parameters, and the loop repeats. No resetting simulations are performed during training. In the SM (Fig.~\ref{fig:sm-pv}), we show the agreement between the predictions and brute-force simulations with resetting for the double-well-to-transition-state transformation. Once the adaptive resetting protocol is trained, we can compare it to other protocols.

Figure~\ref{fig:three} compares the potential switch, a representative trained adaptive resetting policy (weights in Sec.~\ref{sm:training}), and the hybrid protocol.
Resetting reaches the coarse accuracy $J=0.025$ faster than the switch by factors of $45$, $19$, and $69$ for compression, transition-state preparation, and bit erasure, respectively.
However, pure resetting approaches a nonequilibrium steady state (NESS) that deviates slightly from the target and requires continuous entropy production to sustain.
The hybrid protocol overcomes these limitations by switching to the target potential and allowing thermal relaxation to complete the transition.
For the representative policies in Fig.~\ref{fig:three}, it reaches the tight accuracy $(J=10^{-3})$ $2.0$--$2.6\times$ faster than the potential switch.
The hybrid protocol also reduces total entropy production for compression and bit erasure, but incurs a higher cost for transition-state preparation.

We next ask when the speedup from resetting also reduces the thermodynamic cost below the bound for purely diffusive driving, Eq.~\eqref{eq:drivingbound}.
Figure~\ref{fig:bounds} maps the normalized cost $\Sigma D\tau/W_2^2(\rho_0,\rho_\tau)$ across the policy optimization weights and handoff times.
All three transformations cross this bound, but to very different extents.
Bit erasure reaches a minimum of $0.28$ of the bound, with bound crossing over a broad parameter range.
Compression reaches $0.56$ and generally favors an earlier handoff.
Transition-state preparation reaches $0.83$, but only within a narrow parameter range.

These differences reflect two complementary mechanisms of adaptive resetting: nonlocal probability redistribution and selective termination of unfavorable trajectories.
In bit erasure, both mechanisms cooperate: resetting removes particles from the unwanted well and can reinject them into the target well, bypassing slow diffusion across the barrier.
In compression, resetting mainly suppresses excursions outside the target region, though prolonged reinjection can repopulate the tails.
In transition-state preparation, direct reinjection near the barrier is rare, so resetting primarily selects trajectories that reach the barrier by diffusion.
The handoff balances the benefits of continued resetting
against its thermodynamic cost. As resetting approaches
its NESS, further gains in accuracy diminish while
entropy production continues, making thermal relaxation
in the target potential increasingly favorable.

Crossing the purely diffusive bound raises a fundamental
question: what thermodynamic speed limit governs coupled
diffusion--reset dynamics?
We derive a bound for the combined process
(End Matter),
\begin{equation}
\Sigma \;\ge\;
\frac{W_1^2\!\left(\rho_0,\rho_\tau\right)}
{\tau\,\bigl(D+\overline{M}\bigr)}.
\label{eq:floor}
\end{equation}
Here $\overline M$ is the time-averaged mobility of the
reset channel,
$\overline M=\tau^{-1}\int_0^\tau M_{\rm jump}(t)\,dt$.
The instantaneous mobility
$M_{\rm jump}(t)=\frac12\iint m(x,y)(x-y)^2\,dx\,dy$
weights the squared jump distances by the logarithmic
mean $m(x,y)$ of the forward and reverse reset fluxes
(see the End Matter), capturing both the activity and spatial
extent of nonlocal transport.

To derive the bound, we express $W_1$ through
Kantorovich--Rubinstein duality as a supremum over
1-Lipschitz observables. Applying the Cauchy--Schwarz
inequality to the combined contributions of diffusion
and resetting yields the effective mobility
$D+\overline M$.
For diffusion alone, this construction gives a valid
$W_1$ speed limit, weaker than the optimal $W_2$
Langevin bound~\cite{vanvu2023}.
For resetting, the same Lipschitz constraint bounds
the change across a jump $x\to y$ by $|x-y|$,
consistent with $W_1$ speed limits for Markov jump
processes~\cite{vanvu2023,shiraishi2024,dechant2026}.
Combining the two channels gives the mobility
$D+\overline M$.

Unlike the purely diffusive bound, the new limit
depends on the resetting protocol through its
realized mobility $\overline M$.
For the best-performing protocols, the entropy
production exceeds this bound by factors of $3.5$
for erasure, $20$ for compression, and $83$ for
transition-state preparation.
The bound is conservative, in part because the
common $W_1$ construction is weaker than the optimal
$W_2$ bound even for diffusion alone.

In conclusion, we have shown that changing the available transport channels
changes the thermodynamic limits of state preparation.
Adaptive resetting supplements continuous diffusive transport
with nonlocal jumps, enabling rapid preparation across a broad
class of target states.
Combined with final thermal relaxation, the hybrid protocol
crosses the bound for purely diffusive driving in all three
transformations for suitable loss weights and handoff times.

The resetting policy $p_\theta(x)$ is learned from trajectories
without resetting, using a loss function that balances accuracy,
preparation time, and entropy production.
The method requires only samples of the initial distribution
and reweighting of reset-free trajectories, providing a route
to higher-dimensional systems and experimental data.
Here, we assumed instantaneous reset jumps and neglected the
implementation-dependent work of the physical reset mechanism.
Treating finite-time and physically costed
resetting~\cite{olsen2024,gupta2025dynamic,gupta2025erasure}
is a natural next step.

We also derived a thermodynamic speed limit for coupled
diffusion--reset dynamics that accounts for the additional
mobility of nonlocal transport.
Our results establish that combining transport mechanisms opens new pathways for state-to-state transitions, enabling faster transformations without violating thermodynamic speed limits and potentially surpassing the bounds imposed by individual transport mechanisms.

\begin{acknowledgments}
Y.R. and R.G. acknowledge support from the Israel Science Foundation (grants No. 385/21 and 2013/25) and from the European Research Council (ERC) under the European Union’s Horizon 2020 research and innovation program (Grant Agreement No. 101002392).
B.H. acknowledges support from the Israel Science Foundation (grant No. 2572/26).
This work was supported in part by the Zuckerman STEM Leadership Program to D.B.K.C.
The authors acknowledge the use of OpenAI tools (GPT-5.6 and GPT-6) and Anthropic Claude (Opus 4.8, Opus 5, and Fable 5.1) to help edit the manuscript, organize the literature, develop code, run simulations, check derivations, and analyze numerical results. All AI-assisted work was guided by the authors and critically reviewed, checked, and revised by the authors, who take full responsibility for the results and conclusions.
\end{acknowledgments}
\bibliography{cost_accounting}
\nocite{benamou2000}

\onecolumngrid
\vspace{1.5em}
\begin{center}\textbf{END MATTER}\end{center}
\vspace{0.5em}
\twocolumngrid

\appendix

\section{The coupled diffusion--reset bound}
\label{em:floor}
We now derive Eq.~\eqref{eq:floor} for the coupled diffusion--reset dynamics. The two transport channels act simultaneously on the same evolving density, so their separate endpoint speed limits cannot simply be added. Instead, we bound the change of a common 1-Lipschitz observable under both channels. For the diffusive contribution this gives the $W_1$ form of the Langevin transport bound, which is valid but weaker than the optimal $W_2$ bound~\cite{vanvu2023}. For the jump contribution, the same Lipschitz constraint relates the change of the observable to the spatial length of each jump, consistent with $W_1$ speed limits for Markov jump processes~\cite{shiraishi2024}. This common formulation yields a single bound for the hybrid dynamics.

We start from the Fokker--Planck equation under adaptive resetting,
\begin{equation}
\partial_t\rho = \mathcal{L}_{\rm FP}\rho + R(t)\,\rho_R(x) - r(x)\rho(x) ,
\label{eq:em-master}
\end{equation}
with $\mathcal{L}_{\rm FP}$ the Fokker--Planck generator in $U_0$ and
$R(t)=\int r\rho\,dx$ the total reset rate. Then, we treat resetting as a jump network
that carries the flux $q(x,y)=r(x)\rho(x)\rho_R(y)$ from $x$ to $y$. As we show in Sec.~\ref{sm:ep} of the SM, the
Schnakenberg entropy-production rate of that network is
\begin{equation}
\dot\Sigma_{\rm jump}=\tfrac12\!\iint\!\big[q(x,y)-q(y,x)\big]
\ln\frac{q(x,y)}{q(y,x)}\,dx\,dy .
\label{eq:em-jump}
\end{equation}

Using the logarithmic mean $m(a,b):=\tfrac{a-b}{\ln(a/b)}>0$, Eq.~\eqref{eq:em-jump} is exactly a quadratic form in the net edge current $\delta\!j(x,y):=q(x,y)-q(y,x)$:
\begin{equation}
\dot\Sigma_{\rm jump}=\tfrac12\!\iint
\frac{\delta\!j(x,y)^2}{m(x,y)}\,dx\,dy,
\label{eq:em-logmean}
\end{equation}
with $m(x,y) = [q(x,y) - q(y,x)]/\ln[q(x,y)/q(y,x)]$ the logarithmic mean of the two edge fluxes, the same structure as the reaction channel in geometric thermodynamics of reaction--diffusion systems~\cite{nagayama2023}.

We define the jump mobility
\begin{equation}
\begin{split}
&M_{\rm jump}(t):=\tfrac12\!\iint m(x,y)\,(x-y)^2\,dx\,dy\\
&\quad\le\tfrac14\!\iint\big[q(x,y)+q(y,x)\big]\,(x-y)^2\,dx\,dy,
\end{split}
\label{eq:em-mob}
\end{equation}
where the inequality is the log-mean--arithmetic-mean bound. Its right-hand side is half the reset-event rate times the mean-square teleport length.

To derive the bound we start from Kantorovich--Rubinstein duality~\cite{shiraishi2024}, which writes the distance as a supremum over 1-Lipschitz test functions $\phi$ and hence as a time integral along the path,
\begin{equation}
\begin{split}
W_1(\rho_0,\rho_\tau)&=\sup_{\phi}\int\phi\,(\rho_\tau-\rho_0)\,dx\\
&=\sup_{\phi}\int_0^\tau\partial_t\langle\phi\rangle\,dt .
\end{split}
\label{eq:em-kr}
\end{equation}
When we differentiate along the path and insert the master equation~\eqref{eq:em-master}, the transport splits into a diffusive and a jump channel,
\begin{equation}
\partial_t\langle\phi\rangle=\int\phi'\,j\,dx+\tfrac12\!\iint\delta\!j\,\Delta\phi\,dx\,dy ,
\label{eq:em-phidot}
\end{equation}
with $\Delta\phi=\phi(y)-\phi(x)$ the increment of the test function across a jump.
The reset terms antisymmetrize to the jump term, $\int\phi\,(R\,\rho_R-r\rho)\,dx=\tfrac12\iint\delta\!j\,\Delta\phi$, and the boundary term of the diffusive part vanishes for confining $U_0$. Each channel obeys Cauchy--Schwarz separately,
\begin{equation}
\Big|\!\int\!\phi'j\,dx\Big|
\le\sqrt{\dot\Sigma_{\rm diff}\,{\textstyle\int}D\rho\,\phi'^2dx}
\le\sqrt{\dot\Sigma_{\rm diff}\,D} ,
\label{eq:em-csdiff}
\end{equation}
\begin{equation}
\begin{split}
\Big|\tfrac12\!\iint\!\delta\!j\,\Delta\phi\Big|
&\le\sqrt{\dot\Sigma_{\rm jump}\,\tfrac12{\textstyle\iint}m\,\Delta\phi^2}\\
&\le\sqrt{\dot\Sigma_{\rm jump}M_{\rm jump}} ,
\end{split}
\label{eq:em-csjump}
\end{equation}
where we used $|\phi'|\le1$ and $|\Delta\phi|\le|x-y|$. Summing Eqs.~\eqref{eq:em-csdiff} and \eqref{eq:em-csjump} with $(\sqrt{aA}+\sqrt{bB})^2\le(a{+}b)(A{+}B)$ gives
\begin{equation}
\big|\partial_t\langle\phi\rangle\big|\le\sqrt{\dot\Sigma(t)\,\big(D+M_{\rm jump}(t)\big)} .
\label{eq:em-cs1}
\end{equation}
Integrating in time and taking the supremum over $\phi$ in Eq.~\eqref{eq:em-kr} bounds the distance,
\begin{equation}
W_1(\rho_0,\rho_\tau)\le\int_0^\tau\sqrt{\dot\Sigma(t)\,\big(D+M_{\rm jump}(t)\big)}\,dt .
\label{eq:em-cs2}
\end{equation}
One more Cauchy--Schwarz inequality, now in time, separates the entropy production from the mobility,
\begin{equation}
\begin{split}
W_1(\rho_0,\rho_\tau)&\le\sqrt{\int_0^\tau\dot\Sigma\,dt}\;\sqrt{\int_0^\tau\big(D+M_{\rm jump}\big)\,dt}\\
&=\sqrt{\Sigma_\tau\int_0^\tau\big(D+M_{\rm jump}\big)\,dt} ,
\end{split}
\label{eq:em-cs3}
\end{equation}
and squaring gives the bound,
\begin{equation}
\begin{split}
\Sigma_\tau&\;\ge\;\frac{W_1^2(\rho_0,\rho_\tau)}{\tau\,\big(D+\overline{M}\big)},\\
\overline{M}&\;=\;\frac1\tau\int_0^\tau M_{\rm jump}(t)\,dt .
\end{split}
\label{eq:em-floor}
\end{equation}

For diffusion alone, $M_{\rm jump}=0$, and Eq.~\eqref{eq:em-floor}  gives
\begin{equation}
\Sigma_\tau\ge\frac{W_1^2(\rho_0,\rho_\tau)}{D\tau}.
\end{equation}
This is a valid but generally weaker form of the Langevin speed limit, of Eq.~\eqref{eq:drivingbound}. Van Vu and Saito obtain the same continuous $W_1$ formulation from the probability current and show that the tighter optimal-transport result for overdamped Langevin dynamics is instead based on $W_2$~\cite{vanvu2023}. Since $W_1\le W_2$ for probability distributions with finite second moment, the present construction deliberately sacrifices the tighter bound of the purely diffusive sector in order to use one class of test functions for both transport channels. For the nonlocal jump sector, the 1-Lipschitz condition gives $|\phi(y)-\phi(x)|\le|y-x|$ and leads directly to the mobility $M_{\rm jump}$; related $W_1$ speed limits for Markov jump processes were derived in Refs.~\cite{vanvu2023,shiraishi2024}. Equation~\eqref{eq:em-floor} is therefore a common speed limit for the coupled dynamics, not a sum of separate endpoint bounds for diffusion and resetting. It is not claimed to be the optimal transport geometry of the hybrid process. Optimal-transport structures for jump processes have recently been studied separately~\cite{dechant2026}.

Equation~\eqref{eq:em-floor} has to be stated at fixed mobility. Without a cap, the same evolution can be had at arbitrarily small entropy production, so no bound in terms of the endpoint densities alone can hold. Suppose we want the ensemble to follow a given path $\rho(t)$. Equation~\eqref{eq:em-master} fixes only the net reset term that path requires, the difference $s=R\,\rho_R-r\rho$ between reinjection and removal, whose value the path sets to $s=\partial_t\rho-\mathcal{L}_{\rm FP}\rho$. Many policies produce the same difference. Reinjecting at a high total rate $R$ while removing at the matching rate $r_R=[R\,\rho_R-s]/\rho$ leaves $s$ unchanged and drives the flux on every edge up without bound. The two directions then almost balance, $q(x,y)\to R\,\rho_R(x)\rho_R(y)$, a detailed-balanced exchange with the reset distribution, so the jump entropy production falls off as $1/R$ while mass is still teleported. Numerically, $R\,\dot\Sigma_{\rm jump}$ is constant to three digits over five decades of $R$. Such a policy is not free. Equation~\eqref{eq:ep} prices the entropy production of the reset events, but not the work of operating the mechanism that draws the samples, which depends on how the reset is implemented and diverges as it is made sharp and instantaneous~\cite{olsen2024}. Related optimization of resetting under explicit cost functions has been studied in other resetting settings~\cite{gupta2025dynamic}.

\clearpage
\afterpage{\setcounter{page}{1}}
\smfrontmatter
\title{Supplemental Material: Bypassing the Entropy Production Speed Limit of State-to-State Transitions with Adaptive Resetting}

\author{Daniel B. K. Chu}
\affiliation{School of Chemistry, Tel Aviv University, Tel Aviv 6997801, Israel}

\author{R\'emi Goerlich}
\affiliation{School of Chemistry, Tel Aviv University, Tel Aviv 6997801, Israel}

\author{Itamar Steinitz}
\affiliation{School of Chemistry, Tel Aviv University, Tel Aviv 6997801, Israel}

\author{Yael Roichman}
\email{roichman@tauex.tau.ac.il}
\affiliation{School of Chemistry, Tel Aviv University, Tel Aviv 6997801, Israel}
\affiliation{School of Physics, Tel Aviv University, Tel Aviv 6997801, Israel}
\affiliation{The Center for Physics and Chemistry of Living Systems, Tel Aviv University, Tel Aviv 6997801, Israel.}

\author{Barak Hirshberg}
\email{hirshb@tauex.tau.ac.il}
\affiliation{School of Chemistry, Tel Aviv University, Tel Aviv 6997801, Israel}
\affiliation{The Center for Physics and Chemistry of Living Systems, Tel Aviv University, Tel Aviv 6997801, Israel.}
\affiliation{The Center for Computational Molecular and Materials Science, Tel Aviv University, Tel Aviv 6997801, Israel.}

\date{\today}
\maketitle
\renewcommand{\thefigure}{S\arabic{figure}}
\renewcommand{\thetable}{S\arabic{table}}
\renewcommand{\theequation}{S\arabic{equation}}
\setcounter{figure}{0}
\makeatletter
\let\thesubsection\smthesubsection
\let\thesubsubsection\smthesubsubsection
\let\theequation@prefix\smtheeqprefix
\let\@hangfrom@section\smhangfromsection
\let\@sectioncntformat\smsectioncntformat
\@removefromreset{equation}{section}
\makeatother
\setcounter{equation}{0}
\setcounter{table}{0}
\setcounter{section}{0}
\setcounter{secnumdepth}{3}
\renewcommand{\thesection}{S\arabic{section}}

\section{Computational details}
\label{sm:comp}

\subsection{Systems and simulations without resetting}

\emph{Systems.} We work in units $k_BT=\gamma=D=1$ throughout. For harmonic compression, $U_0=x^2/2$ and the target is the equilibrium of $U_{\rm tgt}=k\,x^2/2$ with $k=16$, so the standard deviation shrinks from $1$ to $1/4$. Both double-well transformations start from $U_0=32\,(x^2-x_0^2)^2$ with $x_0=1/2$, a double well with minima at $\pm1/2$ and a barrier of $2\,k_BT$. The transition-state target is the equilibrium of a harmonic well with $k=16$, centered on the barrier top at $x=0$. The erasure target is a harmonic well moment-matched to the conditional $\rho_0(x\,|\,x>0)$, with $k=40.58$ centered at $x=0.434$. 

\emph{Ensemble without resetting.} For each transformation we generate a single ensemble of trajectories without resetting, at the training time step $\Delta t=2.5\times10^{-4}$, integrating the overdamped dynamics with Euler--Maruyama: $800$ trajectories of $2500$ steps for harmonic compression, $6000$ of $600$ steps for double well to transition state, and $250$ of $16000$ steps for bit erasure. 

\subsection{Predictions}

\emph{Predicting from trajectories without resetting.} Given the ensemble without resetting, we predict the behavior under any candidate resetting policy $p_\theta(x)$ without simulating trajectories with resetting. We evaluate the renewal propagator, Eq.~\eqref{eq:renewal} of the main text, as a triangular (Toeplitz) recursion solved once per resetting policy $p_\theta$, over a transient window of $L=600$ steps ($1000$ for bit erasure) on a soft-binned grid ($x\in[-4,4]$, $200$ points, Gaussian kernel of width $3\Delta x$). We describe the training of $p_\theta$ in the next subsection.

\begin{figure}[t]
\includegraphics[width=\columnwidth]
{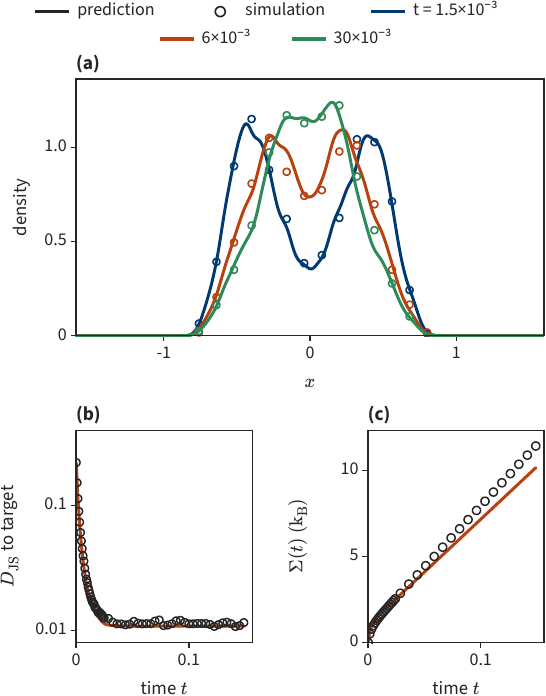}
\caption{For the double-well-to-transition-state resetting policy $p_\theta$: the quantities the loss reads, predicted through the renewal propagator of Eq.~\eqref{eq:renewal} from trajectories without resetting (lines), against direct simulations of trajectories with resetting (circles). (a)~Transient density at $t=1.5,\,6,\,30\times10^{-3}$; (b)~distance to the target $D_{\mathrm{JS}}(t)$; (c)~accumulated entropy production $\Sigma(t)$, against an independent Gaussian-mixture estimator.}
\label{fig:sm-pv}
\end{figure}

\emph{Verifying the predictions.} Figure~\ref{fig:sm-pv} checks these predictions against direct simulations of trajectories with resetting, for the trained resetting policy $p_\theta$ of the double-well-to-transition-state transformation, on every component the loss reads from the propagator. Panel~(a) shows the predicted and simulated densities at three times in the transient. At the training time step, the predicted distance to the target~(b) tracks the simulated one to within $4.4\times10^{-3}$ over the whole transient and settles on the same steady-state floor. The predicted entropy production~(c), Eq.~\eqref{eq:ep} evaluated on the predicted $\rho_n$, agrees to $11\%$ with an independent estimator applied to the simulated ensemble (a Gaussian-mixture fit, then Eq.~\eqref{eq:ep}) at every time beyond the first few steps. The residual comes from the soft binning of the propagator, which underestimates the density tails that the relative-entropy jump term integrates over.

\subsection{Training the neural networks}
\label{sm:training}

The network behind $p_\theta(x)$ has layers $1\to64\to64\to1$ with $\tanh$ hidden activations, initialized to a uniform $p\simeq0.1$ per step. We train with the loss of Eq.~\eqref{eq:loss} of the main text at the accuracy $J=0.025$, using Adam with early stopping. The representative resetting policies of Fig.~\ref{fig:three} use $(\Lambda,w_s,w_c)=(10,30,0.02)$ for harmonic compression and $(10,10,0.05)$ for both double well to transition state and bit erasure. Its hybrid hands off once the reset-only policy reaches the coarse accuracy, at $\tau_{\rm acc}=1.95\times10^{-3}$ for harmonic compression, $2.00\times10^{-3}$ for double well to transition state and $0.65\times10^{-3}$ for bit erasure. Figure~\ref{fig:bounds} is not one policy but a family: we retrain at each weight ratio on a grid of $w_s/w_c$ spanning $0.1$ to $2\times10^{4}$, and deploy each at $22$ handoffs geometrically spaced over $0.081$ to $9.28\,\tau_{\rm acc}$. Figure~\ref{fig:sm-pv} uses a separate double-well-to-transition-state policy, trained at $(10,1.3,1.0)$.

\begin{figure}[t]
\includegraphics[width=\columnwidth]
{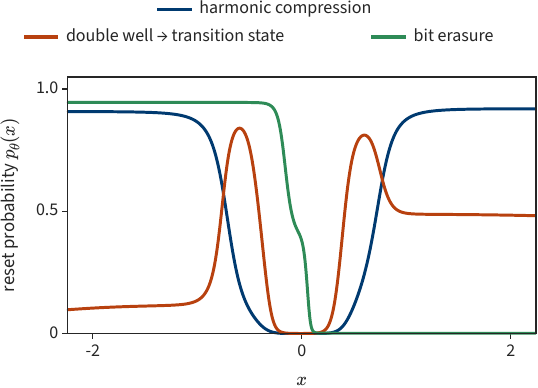}
\caption{Per-step reset probability $p_\theta(x)$ at the training time step for the representative resetting policies of Fig.~\ref{fig:three} of the main text, one curve per transformation: harmonic compression (blue), double well to transition state (orange), and bit erasure (green). Each performs a genuinely different task: the compression policy resets outside the narrow target core, the transition-state policy resets inside the two wells and spares the barrier the target occupies, and the erasure policy resets the entire wrong-well half line, yet all three are learned from the same loss.}
\label{fig:sm-policies}
\end{figure}

Figure~\ref{fig:sm-policies} shows the three trained resetting policies $p_\theta(x)$ used in Fig.~\ref{fig:three} of the main text. Each performs a genuinely different task. For harmonic compression, $p_\theta$ resets everything that strays outside the narrow target core. For double well to transition state, it resets inside the two wells and spares the barrier region the target occupies. For bit erasure, it resets the entire wrong-well half line and leaves the right well alone. All three are learned from the same loss, with no protocol engineering.

The network is not constrained to respect the $x\to-x$ symmetry of the double-well-to-transition-state transformation, so its resetting policy $p_\theta(x)$ is only approximately symmetric. Over the two wells and the barrier top, $p_\theta$ does come out mirror-symmetric, because there the ensemble carries most of its density and the steady state depends sensitively on it. The loss only ever evaluates $p_\theta$ at the positions the ensemble visits. On the steep flanks of the two peaks and in the low-density tails it therefore reads almost no weight, leaves $p_\theta$ free, and the two sides need not mirror. The asymmetry sits where it does not affect the steady state, the speed, or the cost. Imposing the symmetry explicitly (an even parameterization of $p_\theta$) should remove it without changing the result.

\subsection{Simulations with resetting}

A trained resetting policy $p_\theta(x)$ is a per-step probability at the training time step. Resetting relaxes very quickly, and at the training time step the target is reached in only a handful of steps, too few to resolve the time $\tau$ it takes. We therefore train at a step coarse enough that the propagator window spans the whole transient, and run the trajectories with resetting at a finer one, where $\tau$ is converged. To run at another time step, we convert through the $\Delta t$-invariant rate, $r(x)=-\ln[1-p_\theta(x)]/\Delta t_{\rm train}$, and back through $p'=1-e^{-r\,\Delta t}$. The simulations of Figs.~\ref{fig:three} and \ref{fig:bounds} run at the finer $\Delta t=5\times10^{-5}$. Table~\ref{tab:dtconv} shows the convergence for the three representative resetting policies: as $\Delta t$ falls from $5\times10^{-4}$ to $2.5\times10^{-5}$, the time $\tau$ to reach $J=0.025$ plateaus to within a few percent while the step count grows as $1/\Delta t$, so the speedups converge in time rather than in steps.

\begin{table}[htbp]
\caption{Time-step convergence of the three representative resetting policies of Fig.~\ref{fig:three}. For each transformation, the time $\tau$ to reach $J=0.025$ and the number of steps taken, versus $\Delta t$. Every $\tau$ is in units of $10^{-3}$; the first row is the potential switch, for reference. As $\Delta t$ falls, $\tau$ plateaus to within a few percent while the step count grows as $1/\Delta t$, consistent with the speedups quoted in Fig.~\ref{fig:three} of the main text.}
\label{tab:dtconv}
\begin{ruledtabular}
\begin{tabular}{lcccccc}
 & \multicolumn{2}{c}{compression} & \multicolumn{2}{c}{dw$\to$ts} & \multicolumn{2}{c}{erasure}\\
$\Delta t$ & $\tau$ & steps & $\tau$ & steps & $\tau$ & steps\\
\colrule
switch & 87 & & 37 & & 45 & \\
\colrule
$5\times10^{-4}$   & 4.0 & 8  & 4.0 & 8  & 2.5  & 5\\
$2.5\times10^{-4}$ & 2.8 & 11 & 2.8 & 11 & 1.3  & 5\\
$1\times10^{-4}$   & 2.1 & 21 & 2.2 & 22 & 0.8  & 8\\
$5\times10^{-5}$   & 2.0 & 39 & 2.0 & 40 & 0.65 & 13\\
$2.5\times10^{-5}$ & 1.9 & 77 & 1.9 & 77 & 0.63 & 25\\
\end{tabular}
\end{ruledtabular}
\end{table}

We run ensembles of $4\times10^4$ trajectories with resetting and collect them into $800$ bins on $[-4,4]$. The distances plotted in Fig.~\ref{fig:three}(a)--(c) of the main text and in Fig.~\ref{fig:sm-pv}(b) are Jensen--Shannon divergences on that grid.
For display we smooth the density curves in the insets of the same panels, averaging over groups of four neighboring bins. The hybrid protocol resets until the ensemble first reaches $J=0.025$, then evolves in the target potential without resetting, but otherwise uses the same setup.

A finite ensemble sets a sampling floor on the measured divergence. Binning $N$ independent draws from a target on this grid leaves a residual $D_{\rm JS}$ of $4$--$7\times10^{-4}$ at $N=4\times10^4$, so the tight accuracy $J=10^{-3}$ sits only a factor of two above it. Enlarging the ensemble tenfold, to $N=4\times10^5$, lowers the floor by an order of magnitude, well below $J=10^{-3}$, and leaves the hybrid-to-switch speedups nearly unchanged. The $J=10^{-3}$ crossings therefore measure the protocols, not the sampling floor. 

\subsection{Entropy production and bounds}
\label{sm:cost}

\emph{Entropy production.} For the switch and for the hybrid's relaxation phase we accumulate, per step, the heat dissipated into the bath plus the change of the system's Gibbs--Shannon entropy $S=-\int\rho\ln\rho\,dx$ on that grid. The potential is static, so the particle exchanges energy only with the bath: it absorbs $\Delta Q=U(x_{n+1})-U(x_n)$ along a step and dissipates $-\Delta Q$~\cite{smref-sekimoto2010}. We validated this accounting against the exact identity for an instantaneous switch followed by full relaxation, $\Sigma(\infty)=W_{\rm sw}-\Delta F$~\cite{smref-seifert2012}, where $\Sigma(\infty)$ is the total entropy produced once the system has relaxed. Here $W_{\rm sw}=\int\rho_0\,(U_{\rm tgt}-U_0)\,dx$ is the work done by the switch, the mean jump in potential energy at the instant the potential changes at fixed position, and $\Delta F=-\ln(Z_{\rm tgt}/Z_0)$, with $Z=\int e^{-U}dx$, is the free-energy difference between the final and initial equilibria. The two agree to better than $1\%$.

For resetting, we evaluate Eq.~\eqref{eq:ep} of the main text with an analytic Gaussian density (harmonic compression) or a two-component Gaussian mixture with likelihood-based model selection fitted to the ensemble (both double-well transformations), refitted as the ensemble evolves. We cannot use empirical histograms here: the relative-entropy jump term integrates over regions where $\rho_R$ has mass but the ensemble has no counts. Fitting the density also makes the current $j$ and the flux ratio $h$ analytic, so neither the $1/\rho$ of the diffusive term nor $\ln h$ needs regularizing. Equation~\eqref{eq:ep} is a rate, so we evaluate it at every integration step and accumulate $\Sigma(t)=\sum_{n\Delta t\le t}\dot\Sigma(n\Delta t)\,\Delta t$, which is the entropy production plotted in Figs.~\ref{fig:three} and \ref{fig:bounds}. The loss of Eq.~\eqref{eq:loss} instead evaluates Eq.~\eqref{eq:ep} on the soft-binned $\rho_n$, which is differentiable in $\theta$ but underestimates the tails of the jump term; the entropy production we quote and plot in Figs.~\ref{fig:three} and \ref{fig:bounds} comes from the fitted-density path described here, while the soft-binned one is what the propagator predicts, tested in Fig.~\ref{fig:sm-pv}. 

The hybrid uses both in turn. We follow Eq.~\eqref{eq:ep} up to the handoff, the moment the potential is switched, and from the value it has reached there we go on adding the heat and the entropy change of the relaxation.

\emph{Bounds.} Both bounds compare the entropy production with the transport actually performed, so we evaluate the Wasserstein distances between $\rho_0$ and the density \emph{actually reached} at $\tau$, not the idealized target. In one dimension both have closed forms,
\begin{equation}
\begin{split}
W_1(\rho_0,\rho_\tau)&=\int_{-\infty}^{\infty}\big|C_0(x)-C_\tau(x)\big|\,dx,\\
W_2^2(\rho_0,\rho_\tau)&=\int_0^1\big[C_0^{-1}(u)-C_\tau^{-1}(u)\big]^2du,
\end{split}
\label{eq:sm-wass}
\end{equation}
with $C_0$ and $C_\tau$ the cumulative distributions of $\rho_0$ and $\rho_\tau$, and $C^{-1}$ the quantile function, the inverse of $C$. The $W_1$ integral runs over position $x$ and the $W_2$ integral over the cumulative probability $u$, because only for $W_1$ does the quantile integral collapse to the area between the two cumulative distributions. Both are tail-sensitive, so we compute them on wide grids and check that the outermost bins carry negligible mass, since a truncated tail biases them low. 

The resetting bound also needs the jump mobility $M_{\rm jump}$ of Eq.~\eqref{eq:em-mob} and its time average $\overline M$ of Eq.~\eqref{eq:em-floor}. The mobility is a double integral over pairs of positions weighted by the squared jump length, so it needs a better-resolved density than the distances do, bin by bin out into the tails where the $800$-bin grid is empty. We therefore accumulate it from separate runs of the same protocols, with $4\times10^5$ trajectories on a finer grid of $1401$ points on $[-3.5,3.5]$, collected at $13$ times along the transient. Histograms are safe here, unlike for the entropy production: the logarithmic mean $m$ of the two reset fluxes (End Matter of the main text) stays bounded and vanishes wherever either flux does, so empty bins are harmless, whereas $\ln h$, with $h$ the flux ratio of Eq.~\eqref{eq:sm-hdef}, diverges in the sparsely sampled tails. We accumulate $\overline M$ for each deployed protocol.

\section{Entropy production of adaptive resetting}
\label{sm:ep}

\subsection{Derivation of Eq.~\eqref{eq:ep}}

Under adaptive resetting the density obeys the Fokker--Planck equation with source and sink terms,
\begin{equation}
\partial_t\rho = \mathcal{L}_{\rm FP}\rho + R(t)\,\pR(x) - r(x)\rho(x) .
\label{eq:sm-master}
\end{equation}
Here $\mathcal{L}_{\rm FP}\rho=-\partial_x j$ is the Fokker--Planck generator in $U_0$, $j=-D(\partial_x\rho+\beta\rho\,\partial_xU_0)$ the probability current it transports, and $R(t)=\int r\rho\,dx$ the total reset rate. The last two terms are what resetting adds: the sink $-r(x)\rho$ removes probability wherever the policy fires, and the source $R(t)\pR$ puts it back into $\pR$. Resetting itself is a jump network whose states are the particle's positions, with an edge joining any two of them. The flux from $x$ to $y$ is
\begin{equation}
q(x,y)=r(x)\rho(x)\pR(y),
\label{eq:sm-qflux}
\end{equation}
and the reverse flux $q(y,x)$ is positive wherever $r,\pR>0$, so every edge carries flux in both directions and the standard Schnakenberg edgewise entropy production applies~\cite{smref-schnakenberg1976}. This is different from resetting to a single point which would leave every edge one-directional, since the reverse jump never occurs. The total entropy production of such absolutely irreversible process then diverges~\cite{smref-mori2023}, and the finite cost of a point reset is instead the information-theoretic reset entropy of Ref.~\cite{smref-fuchs2016}.

We now derive the entropy production of the jump network which corresponds to the entropic cost of the resetting process. The Schnakenberg entropy-production rate of such a network is
\begin{equation}
\dot\Sigma_{\rm jump}=\tfrac12\!\iint\!\big[q(x,y)-q(y,x)\big]\ln\frac{q(x,y)}{q(y,x)}\,dx\,dy.
\label{eq:sm-lemma}
\end{equation}
We define
\begin{equation}
h(x):=\frac{r(x)\rho(x)}{\pR(x)},\qquad r,\pR>0\ \text{wherever}\ \rho>0 .
\label{eq:sm-hdef}
\end{equation}
The flux then factorizes,
\begin{equation}
q(x,y)=\pR(x)\pR(y)\,h(x),
\label{eq:sm-qfact}
\end{equation}
so the flux ratio is $h(x)/h(y)$ and Eq.~\eqref{eq:sm-lemma} becomes
\begin{equation}
\begin{split}
\dot\Sigma_{\rm jump}&=\tfrac12\!\iint\pR(x)\pR(y)\\
&\quad\times\big[h(x)-h(y)\big]\big[\ln h(x)-\ln h(y)\big]\,dx\,dy.
\end{split}
\label{eq:sm-hh}
\end{equation}
Expanding the product and using $\int\pR\,dx=1$, the double integral collapses to a covariance,
\begin{equation}
\begin{split}
\dot\Sigma_{\rm jump}&=\E_{\pR}[h\ln h]-\E_{\pR}[h]\,\E_{\pR}[\ln h]\\
&=\Cov_{\pR}(h,\ln h).
\end{split}
\label{eq:sm-cov}
\end{equation}
This is the covariance form of Eq.~\eqref{eq:ep} of the main text, and it holds for \emph{arbitrary} adaptive $r(x)$.
The diffusive channel contributes the standard $\dot\Sigma_{\rm diff}=\int j^2/(D\rho)\,dx$~\cite{smref-seifert2012}, so the total rate is $\dot\Sigma=\dot\Sigma_{\rm diff}+\dot\Sigma_{\rm jump}$, which is Eq.~\eqref{eq:ep} of the main text. 

\subsection{Reduction to the entropy production of maintaining a resetting NESS at constant rate}
\label{Sec:SM_ep}

We show here that, in the case of constant resetting rate $r$, the entropy production rate given by Eq.~\eqref{eq:ep} simplifies to the result of Ref.~\cite{smref-mori2023}.
In this previous work, the entropy production rate is evaluated for stochastic resetting with constant rate $r$ to a distribution $\pR$.
Like in our case, resetting to a distribution rather than to a single point allows the application of Schnakenberg formula.
In variance with our case, the resetting rate $r$ does not depend on position and entropy production is evaluated in steady-state.
The total entropy production $\dot \Sigma$ derived in Ref.~\cite{smref-mori2023} (denoted by $\dot S_{\rm total}$ in Eq.~(14) in Ref~\cite{smref-mori2023}) reads
\begin{equation}
    \dot \Sigma = \int \frac{j^2}{\rho D} dx + r\big[D_{\rm KL}(\rho\|\pR)+D_{\rm KL}(\pR\|\rho)\big]
    \label{eq:LikeMori}
\end{equation}
where the first term is denoted as the \textit{local} entropy production rate $\langle \dot S_{\rm local} \rangle = \int j^2/(\rho D) dx$ (defined in Eq.~(6) in Ref.~\cite{smref-mori2023}).

Our results can be simplified when the resetting rate $r$ is constant by using
\begin{equation}
h(x)=r\,\frac{\rho(x)}{\pR(x)},
\label{eq:sm-hconst}
\end{equation}
which allows to replace the covariance term by the elementary expression of the three moments,
\begin{equation}
\begin{split}
\E_{\pR}[h]&=r,\\
\E_{\pR}[h\ln h]&=r\big[\ln r+D_{\rm KL}(\rho\|\pR)\big],\\
\E_{\pR}[\ln h]&=\ln r-D_{\rm KL}(\pR\|\rho),
\end{split}
\label{eq:sm-moments}
\end{equation}
leading to
\begin{equation}
\Cov_{\pR}(h,\ln h)=r\big[D_{\rm KL}(\rho\|\pR)+D_{\rm KL}(\pR\|\rho)\big],
\label{eq:sm-constr}
\end{equation}
which correspond to the second term in Eq.\eqref{eq:LikeMori} above. Our main result Eq.~\eqref{eq:ep} therefore gives the previous result for constant resetting rate $r$ \cite{smref-mori2023}. %

Reference~\cite{smref-mori2023} splits its total entropy production a second way as well, into the heat exchanged between reset events and a reset term, and that splitting is not the one used here. The heat, $-\beta\!\int j\,\partial_xU_0\,dx$~\cite{smref-seifert2012}, is not the entropy production of the diffusive channel. It falls short of it by the system-entropy contribution $-\int j\,\partial_x\ln\rho\,dx$, and their reset term carries that piece instead. The second splitting therefore moves one contribution across the boundary between the channels and without changing the total entropy production. 

The expressions of Ref.~\cite{smref-mori2023} hold at the steady state. Off steady state, Eq.~\eqref{eq:ep} remains the exact total entropy production of the full diffusion-plus-reset jump process, because each channel contributes its own Schnakenberg term additively, so it is the transient generalization of the steady-state result.

\section{Realizability of a target as a resetting NESS}
\label{sm:realiz}

We now ask when a prescribed target $\rho_{\rm tgt}(x)$ is an \emph{exact} NESS of adaptive resetting. Setting $\rho=\rho_{\rm tgt}$ and $\partial_t\rho_{\rm tgt}=0$ in Eq.~\eqref{eq:sm-master} freezes the total reset rate at a constant $R$ and leaves $0=\mathcal{L}_{\rm FP}\rho_{\rm tgt}+R\pR-r\rho_{\rm tgt}$, so the required rate is
\begin{equation}
r(x)=\frac{\mathcal{L}_{\rm FP}\rho_{\rm tgt}(x)+R\,\pR(x)}{\rho_{\rm tgt}(x)} .
\label{eq:realiz-rate}
\end{equation}
A physical rate must satisfy $r(x)\ge0$ everywhere, so an exact resetting NESS exists only if a finite $R$ makes $\mathcal{L}_{\rm FP}\rho_{\rm tgt}+R\pR\ge0$ for all $x$. The smallest such rate is
\begin{equation}
R_\star=\operatorname*{ess\,sup}_x\left[-\frac{\mathcal{L}_{\rm FP}\rho_{\rm tgt}(x)}{\pR(x)}\right]_+ ,
\label{eq:realiz-Rstar}
\end{equation}
where $\operatorname{ess\,sup}$ denotes the essential supremum, the least bound that holds almost everywhere. The target is realizable if and only if $R_\star<\infty$.

The meaning is direct. $\mathcal{L}_{\rm FP}\rho_{\rm tgt}$ describes how the bare dynamics deforms the target: wherever it is negative the dynamics drains probability from the target, and the re-injection $R\pR$ must refill it. The reset distribution must therefore supply probability everywhere the bare dynamics drains the target, so mere overlap of $\pR$ and $\rho_{\rm tgt}$ is not enough. Reference~\cite{smref-keidar2025} met the same obstruction when shaping a two-dimensional target by hand, where a region reachable only through a high-rate region stayed underpopulated. Equation~\eqref{eq:realiz-Rstar} is its quantitative form.

The three transformations divide on this criterion. For harmonic compression the target is realizable: the ratio in Eq.~\eqref{eq:realiz-Rstar} is largest at the trap center and falls off outwards, so $R_\star$ is finite. For both double-well transformations the reset distribution decays like $e^{-32x^4}$ while the harmonic target decays like $e^{-kx^2/2}$, so the target has far heavier tails. There the quartic dynamics drains the target faster than the exponentially smaller $R\pR$ can refill it, and $R_\star$ diverges. The target is therefore not a resetting NESS at all on the unbounded domain, although it is well approximated over the region that carries essentially all the probability.
Even if the target is realizable, the rate of Eq.~\eqref{eq:realiz-rate} carries the ratio $\pR/\rho_{\rm tgt}$, which runs away wherever the target falls off faster than the reset distribution does. Harmonic compression is such a case: its target is narrower than $\pR$, so the required rate diverges in the tails although $R_\star$ is finite. A sigmoid output layer keeps $p_\theta$ below one and the rate finite, so the policy approximates the required profile rather than reproducing it.

Importantly, none of these issues constrain the hybrid, which is why we use it. Resetting needs only bring the density close enough to the target for a final relaxation in $U_{\rm tgt}$ to finish the preparation, so a target that no rate can hold as a NESS is still within reach. Bit erasure makes the point. Its target is not realizable by resetting, and the hybrid still reaches it to the prescribed accuracy, faster than the switch, and at a cost below the purely diffusive driving bound, as Fig.~\ref{fig:bounds} of the main text shows.

\makeatletter
\let\smautobib\auto@bib@innerbib \let\auto@bib@innerbib\@empty
\let\smbibsetup\@bibsetup
\def\@bibsetup#1{\setlength{\topsep}{\z@}\NATx@bibsetnum{\ref{SMLastBibItem}}}
\makeatother
\smrelabel{LastBibItem}{SMLastBibItem}%
\makeatletter\let\auto@bib@innerbib\smautobib\let\@bibsetup\smbibsetup\makeatother

\end{document}